# Validity of the second law in nonextensive quantum thermodynamics

Sumiyoshi Abe[1] and A. K. Rajagopal[2]

[1]*Institute of Physics, University of Tsukuba, Ibaraki 305-8571, Japan*

[2]*Naval Research Laboratory, Washington, DC 20375-5320, USA*

The second law of thermodynamics in nonextensive statistical mechanics is discussed in the quantum regime. Making use of the convexity property of the generalized relative entropy associated with the Tsallis entropy indexed by *q*, Clausius' inequality is shown to hold in the range $q \in (0, 2]$. This restriction on the range of the entropic index, *q*, is purely quantum mechanical and there exists no upper bound of *q* for validity of the second law in classical theory.

Development of nanotechnology raises a new question of physical importance regarding validity of the ordinary thermodynamic principles, both equilibrium and nonequilibrium, that fundamentally apply to macroscopic objects in the thermodynamics limit. In fact, some surprising properties have been reported on microscopic thermodynamic systems (See an interpretive article [1] and the references therein). To understand thermodynamics of small systems, there are currently two approaches: one modifies the thermodynamic relations by taking into account the surface effects, and the other generalizes Boltzmann-Gibbs statistical mechanics by relaxing the additivity properties of the thermodynamic quantities to include nonextensive features of such systems. The former has been initiated by Hill [2] and further elaborated by himself with Chamberlin [3-5]. The latter is represented by nonextensive statistical mechanics [6-8]. This theory is formulated by making use of Tsallis' nonadditive entropy [9] indexed by $q$, the deviation of which from unity may measure smallness of the object under consideration [10]. The present work aims to contribute to the latter approach.

It is not too much to emphasize that the basic thermodynamic principles have been formulated without recourse to the microscopic physical laws although thermodynamics itself is essentially concerned with the molecular theoretic viewpoint of objective materials. It is our standpoint that statistical mechanics may be modified but thermodynamics should remain unchanged, which in turn implies that, when a generalization of traditional statistical mechanics is considered, it is of crucial importance to establish its consistency with the thermodynamic principles. This is the driving motivation of the present investigation.

The first law of thermodynamics describes conservation of energy and therefore it should be manifestly satisfied by any generalization of statistical mechanics. Also, it is not difficult to make the third law characterizing the completely ordered state hold for any generalized entropies. Recently, it has been shown [11-13] that the modified additivity property, termed *pseudoadditivity*, of the Tsallis entropy is a general one which is consistent with the zeroth law. On the other hand, it is fair to say that the second law has almost never been rigorously examined in the context of nonextensive statistical mechanics.

In this paper, we study validity of the second law of thermodynamics in nonextensive quantum thermodynamics of finite-dimensional systems. In particular, we present a proof of Clausius' inequality by employing the generalized quantum relative entropy referred to as the quantum *q*-relative entropy [14,15] associated with the Tsallis entropy. It turns out that the present discussion also gives an insight into the definition of temperature in nonextensive statistical mechanics. This is in parallel with a recent work [16], where the ordinary relative entropy has been used to elucidate the second law in the quantum regime.

Nonextensive quantum statistical mechanics is formulated based on the Tsallis

entropy

$$S_q[\rho] = \frac{1}{1-q}\left(\text{Tr}\,\rho^q - 1\right), \tag{1}$$

where $q$ is the positive entropic index and $\rho$ is the density matrix. Here and hereafter, Boltzmann's constant is set equal to unity for the sake of convenience. This quantity is nonadditive, since for the factorized joint density matrix, $\rho_{\text{I,II}} = \rho_{\text{I}} \otimes \rho_{\text{II}}$, of a bipartite system $(\Sigma_{\text{I}}, \Sigma_{\text{II}})$, $S_q^{\text{I,II}}$ yields the pseudoadditivity relation $S_q^{\text{I,II}} = S_q^{\text{I}} + S_q^{\text{II}} + (1-q) S_q^{\text{I}} S_q^{\text{II}}$ (with the notation $S_q^{\text{I}} \equiv S_q[\rho_{\text{I}}]$ and so on). The last term on the right-hand side violates additivity as long as $q \neq 1$. In the limit $q \to 1$, $S_q[\rho]$ converges to the familiar von Neumann entropy, $S[\rho] = -\text{Tr}(\rho \ln \rho)$: $\lim_{q \to 1} S_q[\rho] = S[\rho]$. Under the constraints on the normalization condition, $\text{Tr}\,\rho = 1$, and the generalized internal energy, $U_q = <H>_q = \text{Tr}(\rho^q H)/\text{Tr}\,\rho^q$ with the system Hamiltonian $H$, the Tsallis entropy is found to be optimized by the following state:

$$\tilde{\rho} = \frac{1}{Z_q} e_q(-\beta^*(H - \tilde{U}_q)), \tag{2}$$

$$Z_q = \text{Tr}\, e_q(-\beta^*(H - \tilde{U}_q)). \tag{3}$$

Here, $e_q(x)$ denotes the $q$-exponential function defined by $e_q(x) = (1 + (1-q)x)_+^{1/(1-q)}$

with the notation $(a)_+ \equiv \max\{0, a\}$, $\tilde{U}_q = \text{Tr}(\tilde{\rho}^q H)/\text{Tr}\tilde{\rho}^q$, and $\beta^* = \beta/\text{Tr}\tilde{\rho}^q$ with the Lagrange multiplier $\beta$ associated with the constraint on the generalized internal energy. It can be ascertained by a direct calculation that the relation, $\partial S_q[\tilde{\rho}]/\partial \tilde{U}_q = \beta$, holds and accordingly the thermodynamic Legendre transform structure is kept unchanged.

Before proceeding to the second law, it seems appropriate to formulate the first law of thermodynamics [17,18] to identify the quantity of heat. For this purpose, consider the generalized internal energy, $U_q = \text{Tr}(\rho^q H)/\text{Tr}\rho^q$. We are concerned with small change of this quantity from $\tilde{U}_q$ and therefore $\rho$ and $\tilde{\rho}$ in eq. (2) are close to each other. Taking the variation of $U_q$, we obtain the first law

$$\delta' Q_q = \delta U_q + \delta' W_q, \tag{4}$$

where $\delta' Q_q$ and $\delta' W_q$ are the small changes of the quantity of heat and the work given by

$$\delta' Q_q = \frac{\text{Tr}[\delta \rho^q (H - U_q)]}{\text{Tr}\rho^q}, \tag{5}$$

$$\delta' W_q = -<\delta H>_q = -\frac{\text{Tr}(\rho^q \delta H)}{\text{Tr}\rho^q}, \tag{6}$$

respectively.

Now let us address ourselves to the second law in nonextensive quantum thermodynamics. Our idea is to take advantage of the quantum $q$-relative entropy of $\rho$ with respect to the reference state, $\sigma$, associated with the Tsallis entropy, which is given as follows:

$$K_q[\rho \| \sigma] = \frac{1}{1-q}[1 - \text{Tr}(\rho^q \sigma^{1-q})]. \tag{7}$$

In the limit $q \to 1$, this quantity tends to the ordinary quantum relative entropy, $K[\rho \| \sigma] = \text{Tr}[\rho(\ln \rho - \ln \sigma)]$ if the support of $\sigma$ has to be equal or larger than that of $\rho$ [19]. (This quantity was employed in Ref. [16] to discuss the second law of quantum thermodynamics.) It is known [15] that $K_q[\rho \| \sigma] \geq 0$ and $K_q[\rho \| \sigma] = 0$ if and only if $\rho = \sigma$. Therefore, the quantum $q$-relative entropy can be utilized for comparing two states.

We wish to compare $\rho$ with $\tilde{\rho}$ in eq. (2) with the assumption that they are close to each other. It is immediate to find $K_q[\rho \| \tilde{\rho}]$ to be given by

$$K_q[\rho \| \tilde{\rho}] = \frac{1}{\text{Tr} \tilde{\rho}^q} \left\{ S_q[\tilde{\rho}] - S_q[\rho] + \beta^* \text{Tr}[\rho^q (H - \tilde{U}_q)] \right\}, \tag{8}$$

where the identical relation, $(Z_q)^{1-q} = \text{Tr} \tilde{\rho}^q$, has been used. Taking the variation of $K_q[\rho\|\tilde{\rho}]$ with respect to $\rho$, i.e., $\rho \to \rho + \delta\rho$ and $\text{Tr}\,\delta\rho = 0$, with fixed $\tilde{\rho}$, we obtain

$$(\text{Tr}\,\tilde{\rho}^q)\,\delta K_q[\rho\|\tilde{\rho}] = -\delta S_q[\rho] + \beta^*\,\text{Tr}[\delta\rho^q(H - U_q)], \qquad (9)$$

where $\delta\rho^q \tilde{U}_q$ has been replaced by $\delta\rho^q U_q$, since $\rho$ is close to $\tilde{\rho}$ and therefore the difference between these two quantities are of the higher-order infinitesimal. Using eq. (5), we further have

$$(\text{Tr}\,\tilde{\rho}^q)\,\delta K_q[\rho\|\tilde{\rho}] = \beta\,\delta' Q_q - \delta S_q[\rho], \qquad (10)$$

or equivalently

$$\delta K_q[\rho\|\tilde{\rho}] = \beta^*\,\delta' Q_q - \delta S_q^{(R)}[\rho], \qquad (11)$$

where $S_q^{(R)}[\rho] = (1-q)^{-1}\ln\{1 + (1-q)S_q[\rho]\} = (1-q)^{-1}\ln(\text{Tr}\,\rho^q)$ is the Rényi entropy. It is of interest to observe in the above equations that inverse temperature associated with the Tsallis entropy is $\beta$, whereas $\beta^*$ for the Rényi entropy [20,21]. In this respect, however, it should be noticed that the Tsallis entropy is stable [22], whereas the Rényi entropy is not [23] and therefore the Rényi entropy cannot be used for generalizing statistical mechanics..

In order to establish Clausius' inequality

$$\beta \, \delta' Q_q \leq \delta S_q[\rho], \tag{12}$$

it is necessary to show that $\delta K_q[\rho \| \tilde{\rho}]$ is negative. This task is, however, not simple because $\delta \rho$ does not commute with $\rho$, in general. However, as shown below, this task can be achieved only for $q \in (0, 2]$.

To calculate $\delta K_q[\rho \| \tilde{\rho}]$, we represent the variation by a trace-preserving completely positive unital map, $\rho \to \rho + \delta \rho \equiv \Lambda(\rho)$:

$$\Lambda(\rho) = \sum_k V_k \rho V_k^\dagger. \tag{13}$$

$V_k$'s are certain operators satisfying the trace-preserving condition, $\sum_k V_k^\dagger V_k = I$, with the identity operator *I*. The unital condition, $\Lambda(I) = I$, leads to $\sum_k V_k V_k^\dagger = I$. These two conditions are compatible if $V_k$'s are normal, that is, $[V_k, V_k^\dagger] = 0$ for $\forall k$. Since $\tilde{\rho}$ is fixed, $\Lambda(\tilde{\rho}) = \tilde{\rho}$, which is fulfilled if $[V_k, \tilde{\rho}] = 0$. Thus, the variation is understood as follows: $\delta K_q[\rho \| \tilde{\rho}] = K_q[\Lambda(\rho) \| \tilde{\rho}] - K_q[\rho \| \tilde{\rho}]$, which can be thought of as (minus of) entropy production.

Now, let *A* be a positive operator. Then, the function is $f(A) = A^q$ ($q > 0$) is operator monotone, that is, for another positive operator, $B$, such that $B \geq A$, holds $B^q \geq A^q$. A very important point is that if $q > 2$, then this operator function does not possess definite convexity. The function $f(A)$ is operator concave (convex), i.e.,

$f(\lambda A + (1-\lambda) B) \geq (\leq) \lambda f(A) + (1-\lambda) f(B)$ if $q \in (0, 1]$ ($q \in (1, 2]$), where $\lambda \in (0, 1)$ [24]. In other words,

$$F(A) = \begin{cases} A^q & \text{for } q \in (0, 1] \\ -A^q & \text{for } q \in (1, 2]. \end{cases} \quad (14)$$

is operator concave. Then, Ando's theorem states (see Ref. [24]) that

$$F\left(\sum_k V_k A V_k^\dagger\right) \geq \sum_k V_k F(A) V_k^\dagger. \quad (15)$$

Using this theorem, we have

$$\tilde{\rho}^{(1-q)/2} [\Lambda(\rho)]^q \tilde{\rho}^{(1-q)/2} \geq \Lambda(\tilde{\rho}^{(1-q)/2} \rho^q \tilde{\rho}^{(1-q)/2}) \quad (q \in (0, 1]), \quad (16)$$

$$\tilde{\rho}^{(1-q)/2} [\Lambda(\rho)]^q \tilde{\rho}^{(1-q)/2} \leq \Lambda(\tilde{\rho}^{(1-q)/2} \rho^q \tilde{\rho}^{(1-q)/2}) \quad (q \in (1, 2]). \quad (17)$$

Therefore, using the definition in eq. (7), we finally obtain

$$K_q[\Lambda(\rho)\|\tilde{\rho}] \leq K_q[\rho\|\tilde{\rho}] \qquad (q \in (0, 2]) \qquad (18)$$

which establishes Clausius' inequality in eq. (12).

In conclusion, we have shown that the second law of thermodynamics holds in nonextensive quantum thermodynamics with $q \in (0, 2]$. Combining this result with the previous works in the literature, we may now conclude that nonextensive statistical mechanics is fully consistent with the principles of thermodynamics in the quantum regime. It is of great interest to see what happens in quantum systems with $q > 2$, where the second law can be violated, in general. In the classical regime, which corresponds to the situation that all relevant density matrices and observables are simultaneously diagonalized in a common basis, there is no restriction on the range of the entropic index, $q$. It should be emphasized that nonextensive statistical mechanics is an approach to nonequilibrium stationary states of small or complex systems. Therefore, the result deduced above is a statement of entropy production in such circumstances. It is in this context that the significance of the result obtained here is to be understood.

**Acknowledgments**

SA would like to thank Professor I. Ojima for valuable discussions. He also thanks

Professor T. Kunihiro and Yukawa Institute for Theoretical Physics, Kyoto University, for hospitality extended to him. AKR acknowledges the Office of Naval Research for partial support of this work.